\documentclass[12pt]{iopart}

\usepackage{graphicx}
\usepackage{url}
\begin{document}

\title{Towards Photoferroic Materials by Design: Recent Progresses and Perspective}

\author{Ivano E. Castelli}
\address{Department of Energy Conversion and Storage, Technical University of Denmark, DK-2800 Kgs. Lyngby, Denmark}
\ead{ivca@dtu.dk}

\author{Thomas Olsen}
\address{Department of Physics, Technical University of Denmark, DK-2800 Kgs. Lyngby, Denmark}
\ead{tolsen@fysik.dtu.dk}

\author{Yunzhong Chen}
\address{Department of Energy Conversion and Storage, Technical University of Denmark, DK-4000 Roskilde, Denmark}
\ead{yunc@dtu.dk}

\begin{abstract}
The use of photoferroic materials that combine ferroelectric and light harvesting
properties in a photovoltaic device is a promising route to significantly improve the efficiency of solar cells. These
materials do not require the formation of a $p-n$ junction and can produce photovoltages well above the value of the band
gap, because of the spontaneous intrinsic polarization and the
formation of domain walls. In this perspective, we discuss the recent
experimental progresses and challenges for the synthesis of these
materials and the theoretical discovery of novel photoferroic
materials using a high-throughput approach. 
\end{abstract}


\submitto{\it J. Phys. Energy}

\section{Introduction}\label{intro:sec}
\begin{figure}
  \includegraphics[width=\linewidth]{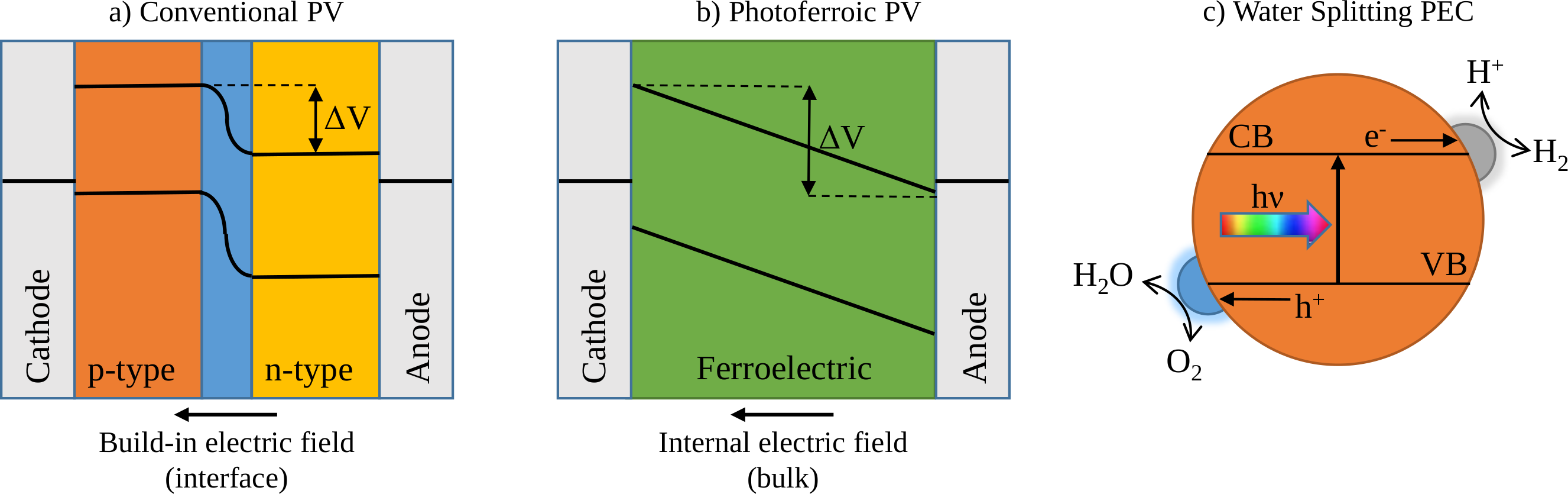}
  \caption{Illustration of conventional PV cell based on a p-n
    junction ($a$), where the splitting of the e-h pair is due to the
    electric field built at the interface, and on photoferroic
    devices, for which the splitting of the photogenerated charges is
    enhanced by the internal polarization ($b$). The scheme of a
    photoelectrochemical cell is also shown ($c$).}
  \label{devices:fig}
\end{figure}

The increase in energy demand and the demand for a society independent from fossil-fuels are making necessary the development of novel sources of
green energy. Photovoltaics (PV) convert solar energy directly into electricity and is one of the dominant technologies
for a green future. In a typical PV device, the solar light is
converted directly into electron-hole pairs, which are separated by a p-n junction and thus result in a voltage difference across the junction (Figure~\ref{devices:fig}a). The electricity
generated can then be connected to the grid or can be stored in
batteries or used to create fuels by means of electrolyzers. Another
technology that relies on harvesting of solar light is a
photoelectrochemical water splitting (PEC) device, in which the
electron-hole pairs are used to split water into hydrogen and oxygen
molecules.\cite{Sivula2016} A fundamental requirement for both technologies is the
efficient absorption of a significant part of the solar spectrum as
well as a slow rate of electron-hole recombination and stability of
the photoactive material. The maximum theoretical efficiency
achievable by a single material in a PV device is around 33 \%, which
corresponds to a band gap of around 1.1 eV, under solar irradiation and
taking into account the possible thermodynamic losses. Many solutions
have been proposed to break this limit, called Schockley-Queisser (SQ)
limit.\cite{Shockley1961} These include the development of cells
based on multi-junctions, solar concentrators, high purity and
thin-film materials. The common requirement of these solutions is that
the maximum photovoltage obtainable by the device correspond to the
sum of the band gaps of the materials involved.\cite{Nayak2019} According to the annual summary of the
progresses of materials and device constructions to reach the highest
efficiencies of PV cells,\cite{NREL_PV} organometal halide perovskites
have recently been suggested as emerging materials for PV devices because of
their good light absorption properties, long lifetime and high mobility of
the photogenerated charges.\cite{PhysRevB.77.045129} However, issues
like limited stability and toxicity (because of
Pb-content) have reduced their applicability in everyday life
devices.\cite{epa_lead}

Another new emerging PV device is constituted by the use of ferroelectric
semiconducting materials, also called photoferroics, as photoabsorbers. This new type of PV device, in principle, can achieve
photovoltages larger than the band gap, thus breaking the SQ
limit. These materials not only absorb sunlight, but also exhibit an intrinsic polarization below a critical temperature (the Curie temperature $T_c$) that can
separate the electron-hole (e-h) pairs without the need of a $p-n$ junction, i.e. these materials can be sandwiched directly between two electrodes. More precisely, the presence of a polar axis allows a second order optical conductivity that can yield a DC current in response to the AC optical frequencies. This leads to
a photovoltaic mechanism very different from the conventional solar cells
based on p-n junctions, as shown in Figure~\ref{devices:fig}b. However,
as we discuss later, there are several limitations in the use of
photoferroic materials in applications and the experimental progresses in the field has so far been rather
limited. A list of the most recent emerging technologies for PV has recently been published.\cite{Wong2019}

Computational methods, in particular in the framework of Density
Functional Theory (DFT),\cite{Hohenberg:1964p3623,Kohn:1965js} have
recently been used to design and discover novel materials with target
properties for several applications, from topological insulators and ferromagnetic materials\cite{Haastrup2018, Olsen2019a, Olsen2019b} to catalysts, batteries and solar energy conversion devices.\cite{Curtarolo2013} Two factors have
contributed to this, namely the increase in the computational power
together with significant methodological improvements, which have made
it possible to study more complex and realistic systems with good
predictability of experimental
results.\cite{Marzari2016,9780080540399,Lejaeghereaad3000} Efficient
light harvesting materials, for example, have been discovered thanks to
high-throughput screening
approaches.\cite{Castelli2012.1,Castelli2012.2,C2EE23482C,Kuhar2017}
In addition, recent progress in first principles discovery of new
ferroelectrics could lead the way to a direct bottom up design of
high-yield photoferroic materials.\cite{Scott2015,Xu2017}

In this perspective, we discuss the current status of photoferroic
materials for solar light conversion and the challenges, in theory as well as experiments, that need to be addressed for a
broader use of this technology. Most of the work on photoferroics is
focused on perovskite materials, because they exist in multiple phases
(cubic, layered, double, ...) and with almost all chemical elements in
the periodic table. They also show a manifold of properties from
ferroelectricity and superconductivity, to efficient light harvesting and high
stability.\cite{Ishihara:2009th} This manuscript is organized as follows: in
Section~\ref{exp:sec}, we report the most recent experimental
progresses; Section~\ref{theo:sec} describes the theoretical
background for photoferroic materials and their in-silico discovery;
and Section~\ref{out:sec} is an outlook on possible routes to solve the identified
challenges.

\section{Recent experimental progresses}\label{exp:sec}
Photoferroic materials combine the properties of a light harvesting
materials and ferroelectrics in a single
compound.\cite{9780306109577} The concept dates back to the late 1970s
and the SbSI is regarded as a prototypical ferroelectric semiconductor
with a band gap of 1.9-2.0 eV and spontaneous polarization of $\sim
25 \mu C/{\rm cm}^2$.\cite{Fong1974} However, a solar cells based on
SbSI was only reported recently with a very promising efficiency of
3-4 \% under solar irradiation.\cite{Nie2017} Photoferroic
solar cells thus remains in its infancy.

Recently, thanks also to the success of halide perovskite solar cells,
photoferroic perovskite materials have gained revived attention in the
community. Perovskite oxides, with a prototypical formula of ABO$_3$, are
the most studied ferroelectric materials and have been used in light
harvesting devices. The large band gap, which is the reason why they
can exhibit large photovoltages, is also the cause of the poor power
conversion efficiency, since only a small fraction of the solar
spectrum is adsorbed and the photocurrent under visible light is small
($\sim$ nA/cm2).\cite{Yuan2014} A good photoferroic material for visible light
absorption should enable the ideal
combination of an optical gap in the range of $1-2$ eV, high electric
polarization, suitable electron-hole mobility, and good
stability.\cite{Song2017} Since 2009, BiFeO$_3$ (BFO)-based materials are
among the most commonly studied photoferroics.\cite{Choi2009} BFO has a
a band gap of around 2.2 eV, but for optimal photoferroic properties it is desirable to find materials with smaller band gaps. The double perovskite
Bi$_2$Fe$_{1-x}$Cr$_x$O$_6$ (BFCO) exhibit a tunable band gap between 1.5 and
2.7 eV, depending on the Fe/Cr ratio and such tunable properites could turn out to be highly promising for the design of novel photoferroic devices.\cite{Nie2017} Moreover, a power
conversion efficiency under AM 1.5 G irradiation ($100 mW/cm^{2}$) of
3.3 \% and a 8.1 \% for Bi$_2$FeCrO$_6$ have been reported for a
single phase solar cell and thin-film solar cells in a multilayer
configuration, respectively.\cite{Nechache2014} The high efficiency of BFCO solar cells was ascribed to the narrow direct band gap, which
enhances the light absorption. First principles calculations have
suggested that the excellent performance also lies in a more efficient
separation of the e-h pairs, which are spatially separated on the Fe
and Cr sites.\cite{Kim2018}

\begin{table}
  \begin{center}
    \caption{Summary of the fundamental properties of the some of the most recently studied
      single layer ferroelectric thin film solar cells: Band gap ($E_g$
      [eV]), Spontaneous polarization ($P$ [$\mu C/{\rm cm}^2$]), Curie
      temperature ($T_c$ [K]), Photocurrent ($J_{sc}$ [$mA/{\rm cm^2}$]),
      Incident wavelength ($\lambda$ [$nm$]), Solar cell Power Conversion Efficiency (PCE) ($\eta$ [\%])}\label{experimental:tab}
    \begin{tabular*}{\textwidth}{@{}l*{9}{@{\extracolsep{5pt}}l}}
      \br Composition&Crystal symmetry&$E_g$ & $P$ & $T_c$ & $J_{sc}$ & $\lambda$ & $\eta$& Refs.\\
      \mr
      BaTiO$_3$ & Tetragonal, P4mm & 3.4 & 29 & 404 & 2.5*10$^{-5}$ & 366 & 1.4*10$^{-4}$ & \cite{Dharmadhikari1982}\\
      PbZr$_{1-x}$Ti$_x$O$_3$ & Cubic, Pm-3m & 3.9 - 4.4 & 82 & 723 & 2.0*10$^{-7}$ & 633 & 1.3*10$^{-7}$& \cite{Dubovik1995}\\
      LiNbO$_3$ & Trigonal, R3c & 3.78 & 72 & 1483 & 1*10$^{-6}$ & 514.5 & 0.03 & \cite{Glass1974} \\
      BiFeO$_3$ & Rhombohedral, R3c & 2.67 & 6.1 (bulk) & 1103 & 1.5 & AM1.5 & 0.17 & \cite{Yang2009}\\
      & & & 80-100 (films) & & & & & \\
      Bi$_2$FeCrO$_6$  & Rhombohedral, R3 & 1.4 - 2.1 & 80 & 485 & 11.7 & AM1.5 & 3.3 & \cite{Nechache2014}\\
      KBiFe$_2$O$_5$ & Orthorhombic, P21cn & 1.6 & 3.73 & 780 & 0.015 & 254 & --- & \cite{Zhang2013}\\
      (K, Ba)(Ni, Nb)O$_{3-\delta}$ & Tetragonal, P4mm & 1.1 - 3.8 & 19 & 673 & 0.004 & 543.5 & --- & \cite{Grinberg2013}\\
      Bi$_x$Mn$_{1-x}$O$_3$ & Monoclinic, C2 & 1.1 & 25 & 770 & 0.25 & AM1.5 & 0.07 & \cite{Chakrabartty2013}\\
      SbSI & Orthorhombic, Pna2 & 2.1 & 20 & 295 & 8.59 & AM1.5 & 3.05 & \cite{Nie2017}\\
      \br
    \end{tabular*}
  \end{center}
\end{table}

Besides BFO-based materials, other
possible photoferroic materials with appropriate band gaps includes
KBiFe$_2$O$_5$,\cite{Zhang2013}
[KNbO$_3$]$_{1-x}$[BaNi$_{0.5}$Nb$_{0.5}$O$_{3-\delta}$]$_x$,\cite{Grinberg2013}
BiFe$_{1-x}$Co$_x$O$_3$,\cite{Machado2019} hexagonal ferrite
(h-RFeO$_3$, R = Y, Dy-Lu) thin films,\cite{Han2018} Ni-doped
SrBi$_2$Nb$_2$O$_9$,\cite{Wu2017} and composite thin films of mixed
BiMnO$_3$ and BiMn$_2$O$_5$.\cite{Chakrabartty2018} Some of the most
promising photoferroics and the typical performance of their single layer ferroelectric thin film solar cells are listed in
Table~\ref{experimental:tab}. Almost all of the substituted single
phase systems end up with either a too large band gap or a too small
polarization, and the power conversion efficiency remains
low. Moreover, oxide perovskites containing 3d transition metal
cations typically have heavy carrier effective masses due to the
non-dispersive band edges derived from the localized $3d$ orbitals. This
results in low mobility and high recombination rates of the carriers.\cite{Wang2016_small} Currently, it still remains unclear whether there is an intrinsic limit
for the hitherto-investigated ferroelectric photoabsorbers in the
power conversion efficiency. On the other hand, several Bi-based double oxoperovskites have shown both narrow band gaps and large charge mobility as well as small exciton binding energies. Ba$_2$Bi$^{3+}$Bi$^{5+}$O$_6$ and Ba$_2$Bi$^{3+}$(Bi$^{5+}_{0.4}$Nb$^{5+}_{0.6}$)O$_6$, for example, have band gaps of 2.1 eV and 1.6 eV,
respectively.\cite{Tang2007} These oxide double
perovskites have shown also high performance as photocatalysts for water oxidation.\cite{Weng2016} Chalcogenide
perovskites have recently been predicted to exhibit narrow band gaps as well as reasonable small
carrier effective masses. Ruddlesden-Popper perovskite sulfides (A$_3$B$_2$S$_7$) could be a new family of
ferroelectric photovoltaic materials with a large photoabsorption of solar light,\cite{Wang2016} as we describe in the next section. However, it is still a challenge to make high
quality sulfides thin films with negligible deep defect states. It has been shown that the large amount of defects and disorder in the bulk structure of the sulfide light-harvesting layer limit the minority charge carrier lifetimes and enhancing recombination processes.\cite{Wallace2017_acs}

\section{In-silico discovery on photoferroic materials}\label{theo:sec}

The discovery of novel materials using quantum mechanical simulations is a
possible way to overcome the challenges described above. In the last
decades, there have been multiple studies on in-silico design of novel
materials for PV devices as well as new ferroelectrics. Novel light
harvesting materials based on inorganic and metal-organic perovskites,
oxynitride and sulfides materials, and known materials from existing
databases are some recent
examples.\cite{Castelli2012.1,Castelli2012.2,C2EE23482C,Kuhar2017,Castelli2013,Castelli2013_DP,Castelli2014_AEM,Castelli2014_APL,Castelli2015,Pilania2016,Kuhar2018,LAgiorgousis2019}
In the case of ferroelectrics there has recently been some progress in obtaining design rules that can readily be applied to computational discovery of new materials exhibiting switchable spontaneous polarization. In particular, for improper ferroelectric perovskites, the spontaneous polarization typically arises from tilts or rotations of the BO$_6$ octahedra\cite{Rondinelli2012,Young2013,Benedek2015} and it has been shown that polar materials can be constructed from inversion symmetric parent compounds by substituting atoms that eliminate inversion centers and conserve certain pseudo inversion centers.\cite{Xu2017}

The combination of optimized absorption properties and ferroelectric order at room temperature define a new route towards the discovery of novel photoferroic materials. In Ref. \cite{Wang2015}, it was shown that the band gap of the ferroelectric KNbO$_3$ could be reduced to the frequency range of visible light by Zn or Bi substitution. Three naturally occurring materials (enargite, stephanite, and bournonite) have recently been suggested as possible candidates from a pool of around 200 materials using criteria based on band gap, optical response, effective masses, and spontaneous polarization.\cite{Wallace2017} Similar criteria have been applied to polar and non-polar chalcogenide perovskites in the layered Ruddlesden-Popper (RP) phase\cite{Wang2016,Zhang2017}: Ca$_3$Z$_2$S$_7$, Ca$_3$Hf$_2$S$_7$, Ca$_3$Zr$_2$Se$_7$, and Ca$_3$Hf$_2$S$_7$ have a good band gaps for PV systems and are stable in a ferroelectric phase with non-trivial polarization. The role of the Goldschmidt tolerance factor\cite{Goldschmidt1926} as a design rule for ferroelectric RP materials has been discussed.
Combining V and Fe in a double perovskite gives rise to a non zero polarization because of a charge transfer from V to Fe, which generates empty V-$d$ orbitals and half-filled Fe-d orbitals. The V-$d^0$ configuration has also the effect of reducing the band gap compared to AVO$_3$ and AFeO$_3$ perovskites, making combinations, like Ba$_2$VFeO$_6$, possible candidates for photoferroic devices.\cite{Chen2017}

\begin{figure}
  \includegraphics[width=\linewidth]{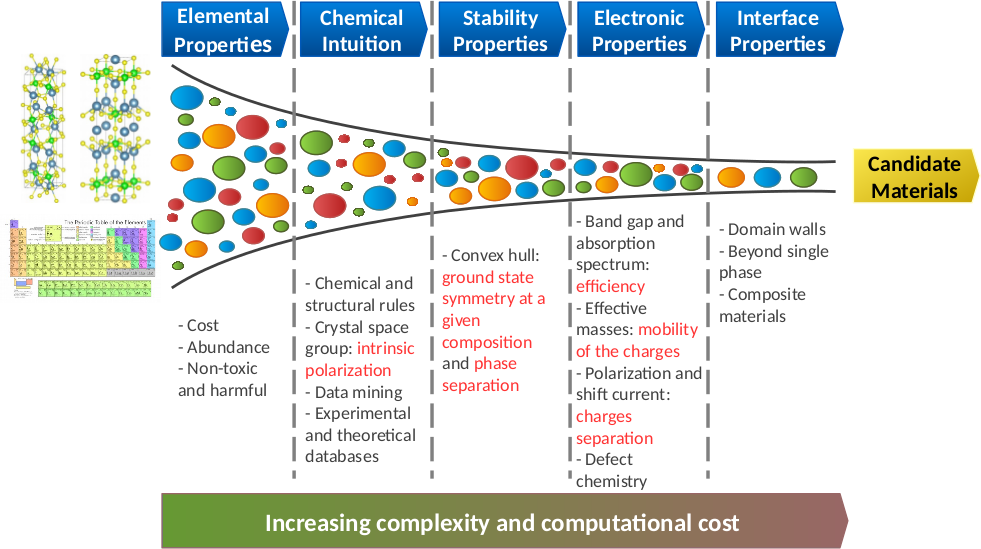}
  \caption{Funnel and required properties for novel photoferroic
    materials.}
  \label{funnel:fig}
\end{figure}

\subsection{High-throughput approach and identification of descriptors}

The materials mentioned above have been identified using high-throughput approaches guided by a screening funnel (Figure~\ref{funnel:fig}). Starting with a large number of possible configurations, at each step of the funnel the computational cost and complexity increases and materials that do not fulfill the required criteria are removed from the pool of possible candidates. The definition of the link (descriptors) between the macroscopic properties required by the device, the selection criteria, and the microscopic quantities calculated with the simulations, is a requirement for any high-throughput screening study.\cite{Curtarolo2013} In the following, we describe the most important descriptors and criteria to identify novel photoferroic materials.

\subsubsection*{Chemical intuition}
Data-mining experimental and computational databases is one of the possible starting points for a screening study. This provides information on possible interesting crystal structures to use for templates or on the chemical elements that have produced interesting materials for the target application. Chemical intuition as well as structural- and chemical-rules,\cite{Jain2013,Castelli2014_rules} such as counting the number of electrons and the tolerance factor, can reduce the number of possible combinations to calculate.

\subsubsection*{Stability}
Stability against phase change and segregation in different materials is a requirement for any novel material. Calculating the stability can be done by comparing the energy of the candidate materials with the energies of the possible competing phases, which are taken from theoretical and experimental databases, like Materials Project,\cite{MaterialsProject} Open Quantum Materials Database (OQMD),\cite{OQMD} or Inorganic Crystal Structure Database (ICSD).\cite{ICSD} At a given composition it is thus possible to identify which structure has the lowest energy, i.e. is stable at 0 K. However, there might be synthesis methods or conditions that allow metastable materials to be synthesized.\cite{Sun2016,Aykol2018} To take this into account, the stability criterion could be relaxed to include a slight metastability (candidate could be considered metastable if their energy is up to 0.2 eV/atom higher than the ground state combination).\cite{Castelli2012.1}

\begin{table}
  \begin{center}
    \caption{Calculated band gap and spontaneous polarization of selected ferrolectric perovskites.}\label{theory:tab}
    \begin{tabular*}{\textwidth}{@{}l*{6}{@{\extracolsep{5pt}}l}}
      \br Material & Crystal symmetry & $E_g$ [eV] & $Pr$ [$\mu$C/cm$^2$] & DFT xc-functional & Refs.\\
      \mr
        BaTiO$_3$ & Tetragonal, P4mm &  1.92 & 26 & LDA & \cite{Piskunov2004,Yuk2017} \\    
        BaTiO$_3$ & Tetragonal, P4mm &  1.84 & 48 & PBE & \cite{Piskunov2004,Yuk2017}\\    
        BaTiO$_3$ & Tetragonal, P4mm &   & 35 & PBEsol & \cite{Yuk2017}\\    
        PbTiO$_3$ & Tetragonal, P4mm & 1.40 & 78 & LDA & \cite{Piskunov2004,Yuk2017}\\    
        PbTiO$_3$ & Tetragonal, P4mm & 1.56 & 128 & PBE & \cite{Piskunov2004,Yuk2017}\\    
        PbTiO$_3$ & Tetragonal, P4mm &   & 98 & PBEsol & \cite{Yuk2017}\\    
        KNbO$_3$ & Tetragonal, P4mm &   & 27 & LDA & \cite{Yuk2017} \\    
        KNbO$_3$ & Tetragonal, P4mm & 1.48 & 51 & PBE & \cite{Faridi2018,Yuk2017}\\    
        KNbO$_3$ & Tetragonal, P4mm &   & 38 & PBEsol & \cite{Yuk2017}\\    
        BiFeO$_3$ & Rhombohedral, R3c & 0.4 & 98.7 & LDA &  \cite{Neaton2005}\\
        BiFeO$_3$ & Rhombohedral, R3c & 1.9 & 93.3 & LDA+U & \cite{Neaton2005}\\
        TbMnO$_3$ & Orthorhombic, Pbnm & 0.5 & -0.47 & LDA+U & \cite{Malashevich2008}\\
        YMnO$_3$ & Hexagonal, P6$_3$cm & 0.45-0.9 & 6.5 & LDA+U & \cite{Biswas2016,Qian2001,Fennie2005}\\
        SrBi$_2$Ta$_2$O$_9$ & Orthorhombic, A2$_1$am & 2.28 & 23.8 & PBE & \cite{Zhao2011,Ke2010}\\    
        Ca$_3$Zr$_2$Se$_7$ & Orthorhombic, Cmc2$_1$& 1.8 & $\sim$16 & HSE06 & \cite{Zhang2017}\\    
        Ca$_3$Hf$_2$Se$_7$ & Orthorhombic, Cmc2$_1$& 2.1 & $\sim$14 & HSE06 & \cite{Zhang2017}\\    
        Ca$_3$Zr$_2$S$_7$ & Orthorhombic, Cmc2$_1$& 2.3 & $\sim$16 & HSE06 & \cite{Zhang2017}\\    
        Ca$_3$Hf$_2$S$_7$ & Orthorhombic, Cmc2$_1$& 2.6 & $\sim$14 & HSE06 & \cite{Zhang2017}\\
      \br
    \end{tabular*}
  \end{center}
\end{table}

\subsubsection*{Spontaneous polarization}
Photoferroic materials rely on the ferroelectricity, so materials, which are not ferroelectric because of their centrosymmetry, could be discarded from the funnel before calculating the electronic properties. Ferroelectric materials are characterized by having a finite electric polarization that can be switched by an external electric field. Such switching involves a structural deformation that connects degenerate ground states with different polarization and a particular ground state will thus have a spontaneously broken symmetry. However, the spontaneous polarization is not well defined for bulk materials, since the integrated dipole density will depend on the choice of unit cell. Polarization changes, on the other hand, are well defined and can be calculated from the polarization current induced by an adiabatic structural deformation. This approach leads to the Berry phase formula,\cite{Xu1993,Vanderbilt1993} which provides a formal expression for the polarization in an atomic configuration parameterized by $\lambda$:
\begin{equation}\label{eq:P}
    \mathbf{P}(\lambda)=\frac{e}{(2\pi)^2}Im\sum_{occ}\int_{BZ}d^3k\langle u_{n\mathbf{k}}|\nabla_\mathrm{k}|u_{n\mathbf{k}}\rangle+\frac{e}{V}\sum_aZ_a\mathbf{R}_a~.
\end{equation}
Here $e$ is the charge of the electron, $u_{nk}$ are the periodic parts of the Bloch states and $R_a$  is the position of atom with nuclear charge $Z_a$. The Berry phase formula can thus be used to calculate changes in polarization under structural deformations and the spontaneous polarization can be defined as the polarization relative to a non-ferroelectric structural phase.
Eq. \ref{eq:P} has been applied to predict the spontaneous polarization of a wide range of ferroelectrics and yield reasonable  agreement with experimental values\cite{Resta} although the results can be somewhat dependent on the choice of exchange-correlation functional.~\cite{Yuk2017} Moreover, the expression provides an easy way to calculate the Born effective charges,\cite{Marzari1998} which determine the atomic displacements under an applied electric field and account for the splitting of longitudinal optical phonons at the Brillouin zone center.\cite{Gonze1997} A selection of ferroelectric materials with calculated polarization and band gaps is displayed in Table~\ref{theory:tab}.

\subsubsection*{Electronic properties}
Once that (meta)stable ferroelectric candidates have been identified, the funnel proceeds with the calculation of the electronic properties. The band gaps (Table~\ref{theory:tab}) obtained from the Kohn-Sham spectrum in DFT are typically significantly smaller than the experimental values. For example, PbTiO$_3$ could appear as being suitable for light harvesting with a gap of $\sim 1.5$ eV, but the experimental gap is more than a factor of two larger, which indicates that it can only absorb a rather small fraction of solar light. The computational discovery of novel photoferroics thus crucially depends on accurate estimates of the band gaps and cannot solely be based on the prediction of local/semi-local functionals like LDA or PBE. The self-interaction error for semi-local exchange-correlation functionals\cite{Perdew1981} and the missing derivative discontinuity\cite{Godby1986} are the main responsible for the underestimation of the band gap. Hybrid functionals, which include a fraction of exact exchange, as for example HSE06,\cite{Heyd2003} many body methods, like the GW approximation,\cite{Aryasetiawan1998} and exchange-correlation functionals, which explicitly include the calculation of the derivative discontinuity, as the GLLB-SC functional\cite{PhysRevB.82.115106} are possible solution to the underestimation of the band gap. The latter, in particular, have been extensively used to identify light harvesting materials using high-throughput approach because of its computational cost, which is comparable with a standard DFT calculation, and accuracy, which is similar to hybrid and many body methods.\cite{Castelli2014_AEM} Although the band gap is the most simple descriptor for the efficiency of a light harvesting device, it gives no information on the character and strength of the transition, which ultimately determines the efficiency. More accurate descriptors that take into consideration these points can be obtained by combining the character of the gap with the shape of the absorption at the band gap edges and the non-radiative recombination losses (spectroscopic limited maximum efficiency (SLME) metric)\cite{Yu2012} or calculating the absorption spectrum by means of time-dependent DFT and convolute it with the solar spectrum.\cite{Castelli2015} Band structures and effective masses can provide useful information as descriptors to the mobility of the electron-hole pair generated by the absorbed photons. The presence of defects might induce mid-gap states in the band gap. Recently a descriptor based on the character of the valence and conduction band to estimate whether a material is defect tolerant or sensitive has been established.\cite{doi:10.1021/jz5001787}

\subsubsection*{Shift current}
The bulk photovoltaic effect denotes the ability of a material to sustain a DC current ($\mathbf{J}$) in response to an AC electric field ($\mathbf{E}$). This is the mechanism underlying photovoltaic response in photoferroics and a quantitative measure of the ability to separate photoexcited electron-hole pairs by intrinsic electric fields in ferroelectrics is provided by the shift current
\begin{equation}
J_i=\sum_{jk}\sigma_{ijk}(\omega)E_j(\omega)E_k(-\omega),
\end{equation}
where subscripts indicate spatial components of current and field.
Clearly, the effect originates from a non-linear response, since the linear response can only give rise to currents oscillating at the same frequency as the perturbing field. Quantitatively the effect is encompassed by the second order conductivity $\sigma_{ijk}$. Since the current as well as the electric field changes sign under space inversion the second order conductivity must vanish in materials with inversion symmetry. Moreover, non-polar materials with inversion symmetry can only sustain a finite shift current under coherent (polarized) illumination and are therefore irrelevant for applications relying on solar light. First principles calculations has established that the shift current accurately accounts for the photocurrent observed in BaTiO$_3$\cite{Young2012} and thus seems to comprise the main mechanism underlying the bulk photovoltaic effect.
Since the shift current mechanism does not rely on a build-in electric field, the photovoltage is not limited by the band gap. The efficiency of photovoltaics based on the bulk photovoltaic effect are thus not limited by the SQ limit and provides an intriguing alternative that may beat the performance of traditional photovoltaics if the right material is discovered.
In Ref. \cite{Tan2016}, the conditions for large non-linear current response was analyzed. Essentially, the response has a structure similar to linear response, except that the dipole transition matrix elements are weighted by a k-point and band dependent “shift vector”. It is thus desirable to have a large joint density of state at the band edges as well as large dipole matrix elements. However, for ferroelectrics the shift vector is not always simply correlated with the spontaneous polarization and the shift current in BaTiO$_3$ and PbTiO$_3$ has similar magnitude despite the fact that the polarization in PbTiO$_3$ is more than twice the value of BaTiO$_3$. In contrast, the shift current in monolayers of Ge and Sn chalcogenides has been shown to be directly proportional to the spontaneous polarization.\cite{Rangel2017} The theoretical prediction of efficient photovoltaics based on the bulk photovoltaic effect thus remains a major challenge and requires full first principles calculations for a given candidate material.
A general ab-initio scheme for calculating the shift current was recently implemented in Wannier90,\cite{IbaezAzpiroz2018} which is interfaced to several electronic structure codes. It is thus straightforward to calculate the shift current in a given material and it seems plausible that high throughput calculations could be based on such approach.

\subsubsection*{Role of domain walls}
Ferroelectric materials under ambient conditions typically exhibit polarization domains, i.e. regions of uniform polarization. The boundaries between domains are known as domain walls and are characterized by a discontinuity in the polarization vectors of adjacent domains. Domain walls thus give rise to local electric fields that may strongly influence the physical properties of ferroelectrics.
In general it is expected that domain walls are detrimental to transport properties since they provide regions of strong scattering. However, it has been demonstrated that domain walls provide a significant contribution to the photovoltaic properties of BiFeO$_3$\cite{Matsuo2016} and for Mn-doped BaTiO$_3$ the bulk photovoltaic effect was reported to increase by orders of magnitude due to domain walls.\cite{Inoue2015} A similar effect has been demonstrated for thin films of BaTiO$_3$.\cite{Zenkevich2014} The reason is likely related to the fact that domain walls or surfaces may provide local electric fields that help separate (first order) photoexcited carriers – similar to the dominating mechanism in traditional p-n photovoltaics. The significance of this effect in photoferroics is still poorly understood and subject to some debate, but for the purpose of comparing calculated photocurrents with experiments, it will be crucial to separate the effect of domain walls and surfaces. This can be done either by comparing shift current calculations to experiments on single domain ferroelectrics or – better yet – to develop a theoretical framework that quantitatively takes the effect of domain walls and surfaces into account. The latter approach could also give rise to new design principles for optimized photoferroics based on domain wall/device architecture.

\subsubsection*{Screening criteria}
The necessary, basic criteria for a materials to be used in a PV device can be summarized as: stability (heat of formation $\le$ 0.2 eV/atom), efficient harvest of visible light (1.0 $\le$ band gap $\le$ 2.0 eV, to account for small inaccuracy in the calculations), and good mobility of the photogenerated charges (small effective masses). These criteria also apply to photoferroic materials, but in that case the presence of a spontaneous polarization and a non-linear photoconductivity comprise additional crucial descriptors for the performance of photoferroic PV devices (the calculation of the shift currents, is still too expensive to be used as descriptor in a screening fashion). These criteria can be applied to various search spaces, both in terms of which chemical elements and crystal structure templates to use. From an academic point of view, the selection of chemical elements should be as broad as possible to avoid ruling out any possible interesting combination. However some chemicals might need to be excluded during the experimental synthesis and device preparation based on considerations such as abundance, toxicity, cost, weight, and so on. Optimal templates for screening projects are structures that can accommodate multiple chemicals and are not too computationally expensive, i.e. have not too many atoms in the unit cell. Some examples are binary structures, like rock-salt and wurtzite, ternary structures, like perovskite or 2D materials, like transition metal dichalcogenides. A detailed description of a high-throughput approach for the discovery of light harvesting materials can be found elsewhere in the literature.\cite{Castelli_book}

\section{Outlook}\label{out:sec}
Single phase photoferroic materials remain rare. The discovery of new ferroelectric materials with a band gap approaching the ideal value of 1.34 eV, corresponding to the maximum efficiency for a single $p-n$ junction PV,\cite{Rhle2016} will definitely represent an important milestone in photoferroelectrics. Alternatively, enhanced photoresponse under visible light of a typical wide-gap ferroelectric materials has also been achieved by combining it with a semiconductor material with a narrow band gap, but the efficiency remains below 1.5 \%.\cite{Zheng2014} So far, the majority of the work has focused on the band-gap engineering of the photoabsorber layer and the carrier mobility remain under-investigated. Besides single phase and composites, double perovskite ferroelectrics with good charge mobility,\cite{Chen2017} can be achieved by layer-by-layer atomically engineered epitaxial oxide interfaces,\cite{Chen2015} and appear highly promising. Furthermore, the optimization of the interfaces or electrodes/photoabsorber connections for ferroelectric PV cell remains rare. Particularly, a Schottky barrier could be formed at the metal-ferroelectric interface of a metal-ferroelectric-metal trilayer solar cell, and the built-in field in the depletion region can largely influence the carrier mobility and may even provide a driving force for the PV effect as in the typical junction-based solar cells. Therefore, new strategies should be set up to optimize the cell efficiency based on photoferroic materials. Additionally, potential improvements in photovoltaic efficiency would also result in greater photocatalytic and photoelectrochemical activity, opening new paths to light energy conversion beyond photovoltaic solar cells. In a photoelectrocatalytic device (PEC), water is dissociated into oxygen and hydrogen molecules by means of solar light (Figure~\ref{devices:fig}c). This technology is based on similar criteria as a PV device plus stability in water environment which can be addressed using Pourbaix diagrams.\cite{PhysRevB.85.235438,Castelli2013_TC} The criterion on the band gap, in addition, should take into account the bare energy to split water is 1.23 eV per molecule and losses like overpotentials associated with the hydrogen and oxygen evolution processes (around 0.1 and 0.4 eV, respectively\cite{Suffredini2000}) and the non-equilibrium condition of the photogenerated electrons and holes has the effect of splitting the Fermi level into electron and hole quasi-Fermi levels (loss of about 0.25 eV for each band edge).\cite{2012} The required band gap should thus be of the order of 2.2 eV, way larger than the required gap for optimal PV materials. The main advantage of PEC over PV is that solar energy is converted into chemical energy (H$_2$ and O$_2$), which are easier to store than the electricity produced by PV devices. Detailed analysis of the feasibility and challenges of PEC devices can be found in the literature.\cite{Sivula2016,Roger2017} The use of photoferroic materials for PEC can improve the efficiency of these devices, by generating large photovoltages, that can more easily provide the required energy to overcome all the losses and overpotentials, while still harvesting a significant fraction of the solar energy. Experiments have already pioneered the use of photoferroics in PEC devices, as highlighted in a recent review.\cite{Kim2018_ap} However most of the materials investigated have a band gap close to the UV part of the spectrum and thus absorbs only a small fraction of the solar energy. The unique combination of ferroelectric and optical properties opens the door to the development of multisource energy harvesting or multifunctional sensing devices for the simultaneous and efficient conversion of solar, thermal, and kinetic energies into electricity in a single material.\cite{Bai2017}

\begin{figure}
  \includegraphics[width=\linewidth]{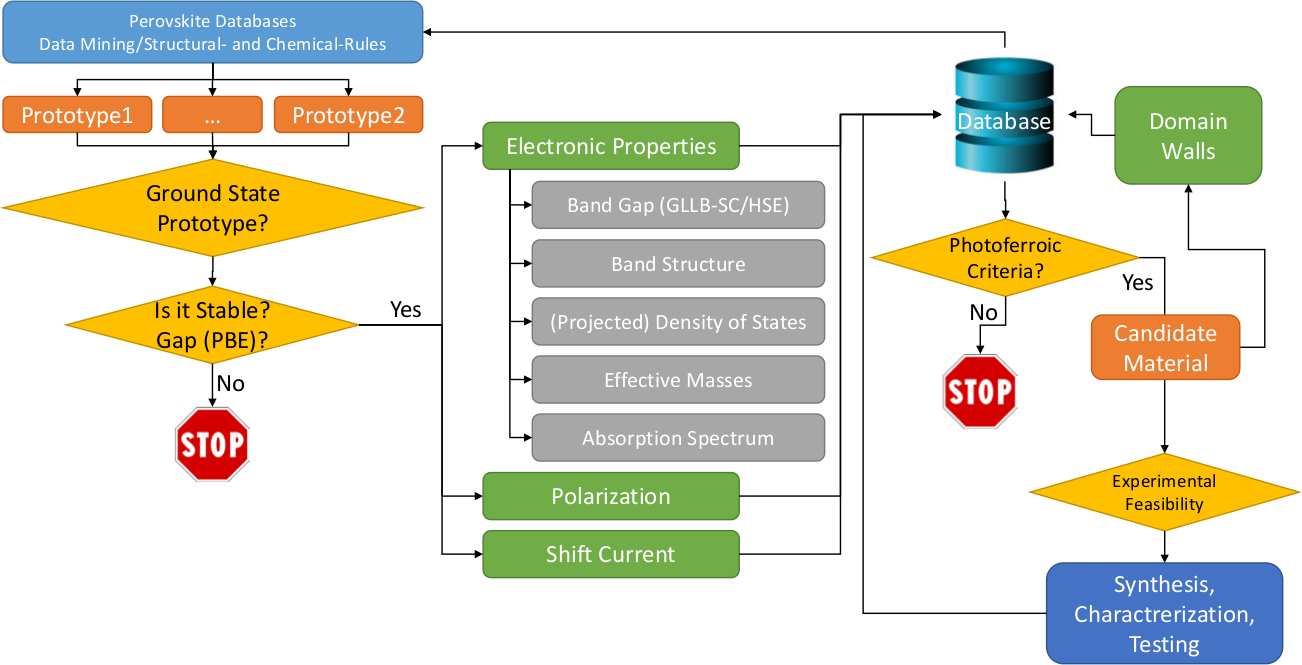}
  \caption{Example of a workflow for the computational discovery of novel photoferroic materials.}
  \label{workflow:fig}
\end{figure}

All these possible applications require a fast discovery of novel materials. Workflows are designed to automate this process and to assure provenance and standardization of the data. Starting from a pool of possible interesting materials, each step of the workflow takes care of a well-defined calculation or decision, which is connected with previous and subsequent steps. As shown in Figure~\ref{workflow:fig}, the candidate materials can be selected from existing databases or using data-mining tools. Different polymorphs or prototypes of the given composition are calculated to identify the ground state structure. Some of these templates are non-polar and will thus not support ferroelectricity and the combination can be discarded. It is important to note that metastable polymorphs can also be interesting and should not be removed from the pool of candidate materials. It has been shown that metastable compounds with an energy of around 100 meV/atom above the ground state can be synthesized and stable at appropriate external conditions.\cite{Sun2016} Once that the ground state is identified, we proceed to calculate the stability against phase separation and the band gap using standard DFT methods at the Generalized Gradient Approximation (GGA) level.\cite{Perdew1992} If the candidate material is (meta)stable and shows a band gap, one proceeds with calculating the electronic properties at a more accurate level as well as polarization and shift current. These data are collected in a database, which will contribute to the selection of the next materials to calculate. Selection criteria will be applied to the candidate materials. If the material fulfills these criteria, it is possible to proceed with an analysis of the experimental feasibility (elemental cost, abundance, toxicity, possible synthesis routes, etc) and eventually with the synthesis and characterization, which will be included in the database to provide the experimental verification. More accurate calculations, for example to understand the role of domain walls, can be performed at this point. Artificial intelligence and machine learning can be helpful in accelerate the materials discovery, although that discussion is beyond the scope of this paper. It is important to note that most of the available workflows are computational and only a few have been developed to cover the full materials discovery from simulations to experiments, especially in connection with organic synthesis and testing.\cite{Roch2018} The complexity of the synthesis for new materials is, in many cases, still a matter of experience and proceeds in a trial-and-error way. A close synergy between experiments and theory, as well as the development of theoretical tools to predict the most promising synthesis path, is one of the ways to solve this issue.

\section{Acknowledgements}
IEC acknowledges support from the Department of Energy Conversion and Storage, Technical University of Denmark, through the Special Competence Initiative "Autonomous Materials Discovery (AiMade)".\cite{aimade} YZC acknowledges the financial support from the European Union's H2020-FETPROACT Grant No. 824072, the support from Agency for Science and Higher Education of Denmark, Grant No. 8073-00014B, and the Independent Research Fund Denmark, Grant No. 9041-00034B.

\section*{References}
\bibliographystyle{iopart-num}
\bibliography{biblio}
\end{document}